\documentclass{article}

\usepackage{PRIMEarxiv}

\usepackage[utf8]{inputenc} 
\usepackage[T1]{fontenc}    
\usepackage{hyperref}       
\usepackage{url}            
\usepackage{booktabs}       
\usepackage{amsfonts}       
\usepackage{amsmath}        
\usepackage{nicefrac}   
\usepackage{float}
\usepackage{microtype}      
\usepackage{lipsum}
\usepackage{fancyhdr}       
\usepackage{graphicx}       
\usepackage{caption}
\usepackage{siunitx}
\usepackage{geometry}
\usepackage{makecell}
\usepackage{subcaption}
\graphicspath{{media/}}     
\usepackage[]{appendix}
\usepackage[backend=biber,style=numeric,sorting=none,natbib=true]{biblatex}
\newcommand{\papertitle}{Physics-Informed Singular-Value Learning for Cross-Covariances Forecasting in Financial Markets}

\title{\papertitle
}

\newcolumntype{T}{S[table-format=1.2(3)e-1]}

\author{
  Efstratios Manolakis \\
  Università di Catania, Dipartimento di Fisica e Astronomia “Ettore Majorana”\\
  Catania, Italy\\
  \texttt{efstratios.manolakis@phd.unict.it} \\
   \And
  Christian Bongiorno\thanks{Corresponding author: \texttt{christian.bongiorno@centralesupelec.fr}}\\
  Université Paris-Saclay, CentraleSupélec, \\Mathématiques et Informatique pour la Complexité et les Systèmes, \\
  91190, Gif-sur-Yvette, France\\
  \texttt{christian.bongiorno@centralesupelec.fr} \\
  \AND
  Rosario N.~Mantegna \\
  Università degli Studi di Palermo, Dipartimento di Fisica e Chimica "Emilio Segrè" \\
  Palermo, Italy\\
  Complexity Science Hub \\
  Vienna, Austria\\
  \texttt{rosario.mantegna@unipa.it} \\
}

\begin{document}
\begin{refsection}

\maketitle

\begin{abstract}
Recent advances in nonlinear shrinkage yield asymptotically optimal cleaners for large covariance matrices and have been extended to empirical cross-covariances via singular-value shrinkage. However, these approaches rely on stationarity and bounded-spectrum assumptions that are violated by real equity returns, which exhibit dependence drift and macroscopic common modes. We propose a physics-informed neural estimator that parameterizes the cleaned cross-covariance matrix in the empirical singular-vector basis and learns a nonlinear map from empirical singular values and marginal projections to cleaned singular values, recovering the cleaning performances of the analytical solution as a limiting case. On U.S. equity data, the learned correction not only improves out-of-sample cross-covariance prediction but also translates these statistical gains into better tracking-error minimization for portfolio replication. Furthermore, it remains stable in regimes where the analytical cross-covariance estimation deteriorates with universe size, suggesting an interpolation between sample-size denoising and a learned forecast correction under non-stationary dependence.
\end{abstract}

\newpage
\section{Introduction}
Empirical cross-covariance matrices summarize linear co-movement between two groups of variables. For zero-mean random vectors $\mathbf{X}\in\mathbb{R}^{n_x}$ and $\mathbf{Y}\in\mathbb{R}^{n_y}$ observed over $\Delta t_\textrm{in}$ paired dates, the cross-covariance matrix $\boldsymbol{\Sigma}_{XY}$ contains the expected products between the two blocks. When the number of variables is comparable to the window length, its sample estimate is strongly affected by noise, which creates spurious relations and weakens Out-Of-Sample (OOS) performance \parencite{chopra1993effect}. Spectral cleaning is well developed for square covariance matrices \parencite{bun2017cleaning}, but a rectangular cross-covariance with $n_x\neq n_y$ requires a Singular Value Decomposition (SVD) and singular-value shrinkage.

Sample cross-covariances are products of observations and therefore involve multiplicative rather than additive noise. Additive-noise results \parencite{gavish2017optimal,troiani2022optimal} do not directly address this setting. Benaych-Georges, Bouchaud, and Potters (BBP) derive an asymptotically optimal rotationally invariant estimator for high-dimensional empirical cross-covariances \parencite{benaych2019optimal}. Their estimator keeps the empirical singular vectors and shrinks the singular values to minimize asymptotic Mean Squared Error (MSE). The solution also depends on projections of the two marginal covariance matrices. However, it is derived under stationarity and bounded-spectrum assumptions, which can fail for equity returns with time-varying dependence and a macroscopic market mode.

Rectangular cross-covariances arise when the two blocks have distinct economic roles. In portfolio replication, a target portfolio on one universe is reproduced with assets from another, and the cross-block covariance enters the tracking-error objective \parencite{cesarone2025benchmark}. Related structures occur when predictors or signals measured on one block are mapped into returns or portfolio directions on another, as in predictive regressions, cross-asset momentum, and principal portfolios \parencite{stambaugh1999predictive,pitkajarvi2020cross,kelly2023principal}. Lead-lag studies provide further examples of dependence across securities and industries \parencite{badrinath1995shepherds,hong2007industries,awijen2023machine}. In our empirical design, however, $X$ and $Y$ are disjoint asset groups observed on the same dates. We estimate their future synchronous rectangular cross-covariance block and use portfolio replication as the downstream financial test; we do not study lead-lag return prediction.

Non-stationarity also changes the interpretation of cleaning. In the covariance setting, stationary shrinkage can underperform simple baselines on real markets \parencite{bongiorno2023filtering}. Physics-informed neural models address this gap by retaining random-matrix structure while adding trainable degrees of freedom \parencite{bongiorno2025neural,bongiorno2025end}. On stationary simulations or shuffled controls, the difference between the in-sample estimate and the OOS target is due to finite-sample noise, so the performed task is denoising. On chronological market data, persistent changes in dependence create an additional forecasting component. A symmetry-preserving model can therefore combine spectral denoising with a correction for the future dependence structure.

Building on BBP theory, we propose a physics-informed neural estimator in the empirical singular-vector basis. It learns a dimension-agnostic map from empirical singular values and marginal projections to cleaned singular values while preserving the two-sided rotational symmetries of the problem. The analytical BBP cleaner is replicated in a limiting stationary case; the neural network is therefore a structured generalization rather than a neural implementation of BBP. The additional flexibility allows the estimator to respond to market modes and persistent dependence changes. We evaluate this behavior on diagnostic synthetic benchmarks, U.S. equity data, and a portfolio-replication problem.

\section{Two-Stream Architecture for Cross--Covariance Cleaning} \label{sec:architecture}
We start from the empirical cross-correlation $\widehat{\mathbf{C}}_{XY}\in\mathbb{R}^{n_x\times n_y}$ computed over $\Delta t_\textrm{in}$ observations, together with the marginal empirical correlations $\widehat{\mathbf{C}}_{XX}$ and $\widehat{\mathbf{C}}_{YY}$. Let $r:=\min(n_x,n_y)$ and write
\begin{equation}\label{eq:svd}
\widehat{\mathbf{C}}_{XY}=\sum_{k=1}^{r}\widehat{s}_k\,\widehat{\mathbf{u}}_k\widehat{\mathbf{v}}_k^\top.
\end{equation}
A Rotationally Invariant Estimator (RIE) keeps the empirical singular vectors and replaces each $\widehat{s}_k$ with a cleaned value $\widetilde{s}_k$. When needed, the singular-vector sets are completed to orthonormal bases; the added directions have zero singular values.

Following the sufficient statistics used by BBP shrinkage \parencite{benaych2019optimal}, we define the marginal projections
\begin{equation}
\widehat{\boldsymbol{\gamma}}^{(x)} := \left(\widehat{\mathbf{u}}_k^{\top}\widehat{\mathbf{C}}_{XX}\widehat{\mathbf{u}}_k\right)_{k=1}^{n_x},
\qquad
\widehat{\boldsymbol{\gamma}}^{(y)} := \left(\widehat{\mathbf{v}}_k^{\top}\widehat{\mathbf{C}}_{YY}\widehat{\mathbf{v}}_k\right)_{k=1}^{n_y},
\end{equation}
and the aspect ratios $q_x:=n_x/\Delta t_\textrm{in}$ and $q_y:=n_y/\Delta t_\textrm{in}$. To retain the marginal directions on the larger side, we set $p:=\max(n_x,n_y)$ and pad the inputs for $k=1,\ldots,p$,
\begin{equation}\label{eq:extention}
\overline{s}_k :=
\begin{cases}
\widehat{s}_k, & k\le r,\\
0, & k>r,
\end{cases}
\quad
\overline{\gamma}^{(x)}_k :=
\begin{cases}
\widehat{\gamma}^{(x)}_k, & k\le n_x,\\
0, & k>n_x,
\end{cases}
\quad
\overline{\gamma}^{(y)}_k :=
\begin{cases}
\widehat{\gamma}^{(y)}_k, & k\le n_y,\\
0, & k>n_y.
\end{cases}
\end{equation}

\begin{figure}[tbh]
    \centering
    \includegraphics[width=0.95\columnwidth]{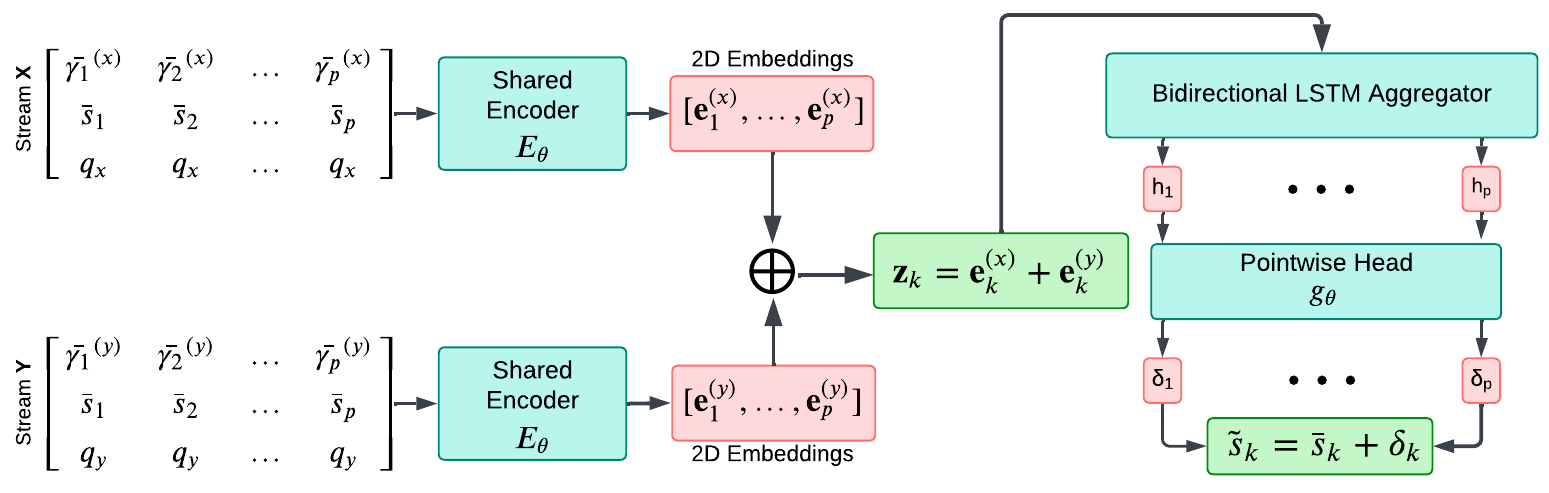}
    \caption{Flowchart of the neural singular value cleaning architecture. The diagram illustrates the construction of dual-stream tokens from marginal projections $\overline{\gamma}$ and singular values $\overline{s}$, their transformation through a shared encoder $E_{\theta}$, and the global context aggregation via a bidirectional LSTM and pointwise head $g_\theta$ to produce the final additive corrections $\delta_k$.}
    \label{fig:Flowchart}
\end{figure}

For each index, the two streams receive
\begin{equation}
\boldsymbol{\tau}^{(x)}_k := \bigl[\overline{\gamma}^{(x)}_k,\ \overline{s}_k,\ q_x\bigr]^{\top},
\qquad
\boldsymbol{\tau}^{(y)}_k := \bigl[\overline{\gamma}^{(y)}_k,\ \overline{s}_k,\ q_y\bigr]^{\top}.
\end{equation}
A shared token-wise encoder $E_\theta$ maps them to two-dimensional embeddings,
\begin{equation}
\mathbf{e}^{(x)}_k=E_{\theta}(\boldsymbol{\tau}^{(x)}_k),
\qquad
\mathbf{e}^{(y)}_k=E_{\theta}(\boldsymbol{\tau}^{(y)}_k),
\end{equation}
which are fused as
\begin{equation}
\mathbf{z}_k=\mathbf{e}^{(x)}_k+\mathbf{e}^{(y)}_k,\qquad k=1,\ldots,p.
\end{equation}
The encoder is a two-layer MLP with 16 hidden units. A two-layer bidirectional LSTM with hidden sizes 128 and 64 supplies global spectral context, and a pointwise two-layer MLP with 252 hidden units returns $\delta_k$. All MLP hidden layers use LeakyReLU activation. The cleaned singular values are
\begin{equation}\label{eq:s_cleaned}
\widetilde{s}_k=\widehat{s}_k+\delta_k,\qquad k=1,\ldots,r.
\end{equation}
\ref{app:BBP_as_limit} shows that this class contains the BBP cleaner. Thus, the network does not implement BBP directly: it preserves the empirical singular-vector basis, invariances, and spectral inputs, but learns a broader mapping when the analytical assumptions are not satisfied. A bounded multiplicative alternative is discussed in \ref{app:FeasibleSV}. Figure~\ref{fig:Flowchart} summarizes the data flow.

We reconstruct the cleaned cross-covariance as
\begin{equation}
\widetilde{\boldsymbol{\Sigma}}_{XY}=\widehat{\boldsymbol{D}}_X\widetilde{\mathbf{C}}_{XY}\widehat{\boldsymbol{D}}_Y.
\end{equation}
where $\widehat{\boldsymbol{D}}_X$ and $\widehat{\boldsymbol{D}}_Y$ are diagonal matrices of in-sample sample standard deviations. The model is trained in the cross-correlation domain; including this rescaling in the objective gives the corresponding end-to-end cross-covariance formulation. For the matrix-prediction experiments, we minimize the BBP-aligned OOS Frobenius loss
\begin{equation}
\mathcal{L}(\theta)=\frac{1}{n_x n_y}\,\bigl\|\widetilde{\mathbf{C}}_{XY}-\mathbf{C}^{\mathrm{OOS}}_{XY}\bigr\|_F^2.
\end{equation}
This loss is task-specific rather than architectural; Section~\ref{sec:tracking_error} uses the same module within a tracking-error objective. The model has $331{,}355$ parameters and is dimension-agnostic and permutation-equivariant. An open-source TensorFlow/Keras implementation of the rectangular cross-correlation estimator is provided by the CRIENet package \cite{manolakis2026crienet}.

\section{Validation on Synthetic and Market Data}\label{sec:validation}
We first benchmark our Neural-Network (NN) estimator on synthetic datasets whose purpose is diagnostic rather than to provide a complete distributional model of equity returns. The benchmarks are designed to separate the main mechanisms relevant to the proposed estimator: finite-sample denoising, tail thickness, boundedness violations, and conditional dependence dynamics. Panels A-C of Table~\ref{tab:synthetic_benchmarks} consider stationary regimes close to those assumed in the BBP theory, namely finite-rank signals and heavy-tailed noise \parencite{benaych2019optimal}. In these settings, the NN matches the BBP implementation released by \textcite{Opt_clean_BBP}, indicating that the network does not generate an artificial gain when the analytical assumptions are approximately satisfied. The Gaussian rows in Panels B--C are therefore reference cases within the heavy-tailed benchmarks, not assumptions on financial returns. Panel D isolates a different mechanism by injecting a macroscopic one-factor mode into the population correlation $\boldsymbol{C}$, which violates the BBP boundedness condition \parencite{benaych2019optimal} and degrades BBP performance. Finally, to provide a finance-style multifactor benchmark, Panel E considers a conditional factor-DCC-GARCH model in which returns follow synthetic Fama--French-style linear factor structures with $d_f\in\{3,4,5\}$ latent factors, factor volatilities follow GARCH recursions, and factor correlations follow DCC dynamics \parencite{fama2015five,engle2002dynamic}. The NN achieves the lowest mean MSE in all three multifactor conditions. Details on the boundedness issue, the CV procedure, and the data-generating processes and validation protocol are provided in \ref{app:CV}, \ref{app:Synthetic}, and \ref{app:FactorDCC}.

\begin{table}[tbh]

\centering
\providecommand{\sym}[1]{\textsuperscript{#1}}
\footnotesize
\setlength{\tabcolsep}{4pt}
\renewcommand{\arraystretch}{1.05}

\begin{tabular}{@{}lllll@{}}
\toprule
Condition & MLE & BBP & CV & NN \\
\midrule

\multicolumn{5}{c}{\textit{Panel A. Finite-Rank Spiked Models}} \\
\midrule
$\xi=40\%$ & $2.00{\times}10^{-3}$ & $2.46{\times}10^{-4}$\sym{***} & $2.46{\times}10^{-4}$\sym{***} & $2.53{\times}10^{-4}$ \\
$\xi=20\%$ & $2.00{\times}10^{-3}$ & $1.34{\times}10^{-4}$\sym{***} & $1.34{\times}10^{-4}$\sym{***} & $1.34{\times}10^{-4}$\sym{***} \\
$\xi=10\%$ & $2.00{\times}10^{-3}$ & $7.00{\times}10^{-5}$\sym{***} & $7.01{\times}10^{-5}$\sym{***} & $7.10{\times}10^{-5}$ \\
$\xi=0\%$  & $2.00{\times}10^{-3}$ & $7.87{\times}10^{-8}$\sym{***} & $2.67{\times}10^{-7}$ & $5.72{\times}10^{-6}$ \\
\midrule

\multicolumn{5}{c}{\textit{Panel B. Heavy-Tailed Bulk Models}} \\
\midrule
Gaussian      & $2.00{\times}10^{-3}$ & $6.16{\times}10^{-4}$ & $6.18{\times}10^{-4}$ & $6.16{\times}10^{-4}$\sym{***} \\
$\alpha=5.0$ & $2.00{\times}10^{-3}$ & $6.17{\times}10^{-4}$ & $6.18{\times}10^{-4}$ & $6.16{\times}10^{-4}$\sym{***} \\
$\alpha=2.5$ & $2.00{\times}10^{-3}$ & $6.16{\times}10^{-4}$ & $6.18{\times}10^{-4}$ & $6.16{\times}10^{-4}$\sym{***} \\
$\alpha=1.5$ & $2.00{\times}10^{-3}$ & $6.19{\times}10^{-4}$\sym{***} & $6.21{\times}10^{-4}$ & $6.19{\times}10^{-4}$\sym{***} \\
\midrule

\multicolumn{5}{c}{\textit{Panel C. White Heavy-Tailed Models}} \\
\midrule
Gaussian      & $2.00{\times}10^{-3}$ & $5.64{\times}10^{-4}$\sym{***} & $5.65{\times}10^{-4}$ & $5.64{\times}10^{-4}$\sym{***} \\
$\alpha=5.0$ & $2.00{\times}10^{-3}$ & $5.65{\times}10^{-4}$\sym{***} & $5.66{\times}10^{-4}$ & $5.65{\times}10^{-4}$\sym{***} \\
$\alpha=2.5$ & $2.00{\times}10^{-3}$ & $2.58{\times}10^{-4}$\sym{***} & $2.59{\times}10^{-4}$\sym{***} & $2.59{\times}10^{-4}$\sym{***} \\
$\alpha=1.5$ & $2.00{\times}10^{-3}$ & $4.15{\times}10^{-5}$\sym{***} & $4.13{\times}10^{-5}$\sym{***} & $4.37{\times}10^{-5}$\sym{***} \\
\midrule

\multicolumn{5}{c}{\textit{Panel D. Gaussian Bulk Models with Mode}} \\
\midrule
$m=0.5$  & $1.13{\times}10^{-3}$ & $7.93{\times}10^{-2}$ & $7.92{\times}10^{-4}$\sym{***} & $7.89{\times}10^{-4}$\sym{***} \\
$m=0.3$  & $1.67{\times}10^{-3}$ & $9.95{\times}10^{-3}$ & $9.94{\times}10^{-4}$\sym{***} & $9.92{\times}10^{-4}$\sym{***} \\
$m=0.1$  & $1.96{\times}10^{-3}$ & $8.63{\times}10^{-4}$ & $8.44{\times}10^{-4}$\sym{***} & $8.54{\times}10^{-4}$ \\
$m=0.05$ & $1.99{\times}10^{-3}$ & $7.44{\times}10^{-4}$\sym{***} & $7.46{\times}10^{-4}$ & $7.62{\times}10^{-4}$ \\
\midrule

\multicolumn{5}{c}{\textit{Panel E. FF-Style Factor-DCC-GARCH Models}} \\
\midrule
FF-3 & $4.07{\times}10^{-3}$ & $3.35{\times}10^{-3}$ & $2.91{\times}10^{-3}$ & $2.33{\times}10^{-3}$\sym{***} \\
FF-4 & $4.40{\times}10^{-3}$ & $3.44{\times}10^{-3}$ & $3.28{\times}10^{-3}$ & $2.44{\times}10^{-3}$\sym{***} \\
FF-5 & $4.58{\times}10^{-3}$ & $3.65{\times}10^{-3}$ & $3.49{\times}10^{-3}$ & $2.68{\times}10^{-3}$\sym{***} \\
\bottomrule
\end{tabular}

     \caption{Average MSE across estimators in synthetic regimes over 1{,}000 simulations. Stars identify the estimator, or statistically indistinguishable best group, with the lowest mean loss in each row. Statistical significance is assessed using the paired nonparametric bootstrap
described in \ref{app:bootstrap}, resampling the common Monte Carlo
realizations and using the same resampled indices for all estimators, with loss
differentials computed against all non-starred alternatives:
* $p<0.10$, ** $p<0.05$, and *** $p<0.01$. Panel A varies the spike fraction $\xi$ (finite-rank). Panels B-C vary the tail exponent $\alpha$ (heavy-tailed bulk and white), with smaller $\alpha$ indicating heavier tails. Panel D repeats Panel B (Gaussian) with an injected one-factor mode. Panel E reports FF-style conditional multifactor DCC-GARCH models with $d_f\in\{3,4,5\}$ latent factors  (see \ref{app:Synthetic} for definitions). All simulations use $n_x=200$, $n_y=350$, and $\Delta t_\textrm{in}=500$, following \cite{benaych2019optimal}; Panel E sets $n=550$ and the same $n_x$, $n_y$, and $\Delta t_\textrm{in}$. Abbreviations: NN = Neural Network estimator; MLE = Maximum-Likelihood Estimator; BBP = Benaych-Georges, Bouchaud, and Potters analytical estimator; CV = Monte Carlo Cross-Validated.}
    \label{tab:synthetic_benchmarks}
\end{table}

On U.S. equities, we use daily adjusted-close returns for the top-1000 stocks by market capitalization over 1995-2024. The eligible universe is reconstituted at each window using only information available before the evaluation date. The filtering procedure is designed to obtain an investable and sufficiently liquid common-equity universe: we exclude ETFs, funds, and non-operating vehicles; retain common shares with a stable trading history; impose auction-participation, liquidity, price, and size requirements; keep only one share class per issuer; and remove near-duplicate tickers with very high in-sample correlation. These filters reduce look-ahead bias, illiquidity-driven noise, and mechanical duplication in the cross-sectional universe; the full thresholds are reported in \ref{app:real}.

For each walk-forward split, the model is trained on data available up to the previous year-end and evaluated on the subsequent calendar-year test segment. A training or test instance is generated by drawing an in-sample window of length $\Delta t_{\mathrm{in}}$, selecting $n=n_x+n_y$ stocks from the eligible universe, and randomly partitioning them into two disjoint synchronous blocks $X$ and $Y$. The in-sample object $\widehat{\mathbf C}_{XY}$ is the cross-correlation matrix between the two blocks computed on the same dates; the OOS target $\mathbf C^{\mathrm{OOS}}_{XY}$ is computed for the same two blocks on the immediately subsequent window of length $\Delta t_{\mathrm{out}}=240$ trading days. Thus, the empirical task is not lead-lag return prediction, but the prediction of the future synchronous rectangular cross-correlation block. During training, we randomize $n\sim\mathcal{U}[50,500]$, $\nu:=n_x/(n_x+n_y)\sim\mathcal{U}[0.05,0.95]$, and $\Delta t_{\mathrm{in}}\sim\mathcal{U}[200,1200]$ to avoid specialization to a single dimension, aspect ratio, or lookback length. Full universe filters and optimization details are reported in \ref{app:real}.

\begin{table}[tbh]
    \centering

\centering
\providecommand{\sym}[1]{\textsuperscript{#1}}
\footnotesize
\setlength{\tabcolsep}{2.6pt}
\renewcommand{\arraystretch}{0.95}

\begin{tabular}{@{}lllllll@{}}
\toprule
Condition & MLE & FF3 & FF5 & BBP & CV & NN \\
\midrule

\multicolumn{7}{c}{\textit{Panel A. Chronological Split} $(\nu=0.25)$} \\
\midrule
$n=250$  & $0.0342$ & $0.0341$ & $0.0337$ & $0.0303$\sym{***} & $0.0330$ & $0.0295$\sym{***} \\
$n=500$  & $0.0340$ & $0.0339$ & $0.0335$ & $0.0317$ & $0.0328$ & $0.0277$\sym{***} \\
$n=750$  & $0.0340$ & $0.0339$ & $0.0335$ & $0.0473$ & $0.0328$ & $0.0298$\sym{***} \\
$n=1000$ & $0.0340$ & $0.0339$ & $0.0335$ & $0.0656$ & $0.0328$ & $0.0296$\sym{***} \\
\midrule

\multicolumn{7}{c}{\textit{Panel B. Shuffle Control} $(\nu=0.25)$} \\
\midrule
$n=250$  & $0.0111$ & $0.0124$ & $0.0119$ & $0.0112$ & $0.0106$\sym{***} & $0.0106$\sym{***} \\
$n=500$  & $0.0112$ & $0.0124$ & $0.0119$ & $0.0224$ & $0.0106$\sym{***} & $0.0107$\sym{***} \\
$n=750$  & $0.0112$ & $0.0123$ & $0.0119$ & $0.0456$ & $0.0106$\sym{***} & $0.0107$\sym{***} \\
$n=1000$ & $0.0112$ & $0.0124$ & $0.0119$ & $0.0684$ & $0.0106$\sym{***} & $0.0107$\sym{***} \\
\midrule

\multicolumn{7}{c}{\textit{Panel C. Chronological Split} $(n=1000)$} \\
\midrule
$\nu=0.1$ & $0.0340$ & $0.0339$ & $0.0335$ & $0.0431$ & $0.0328$ & $0.0295$\sym{***} \\
$\nu=0.2$ & $0.0341$ & $0.0340$ & $0.0336$ & $0.0601$ & $0.0329$ & $0.0296$\sym{***} \\
$\nu=0.3$ & $0.0340$ & $0.0339$ & $0.0335$ & $0.0687$ & $0.0328$ & $0.0296$\sym{***} \\
$\nu=0.4$ & $0.0341$ & $0.0339$ & $0.0336$ & $0.0740$ & $0.0328$ & $0.0297$\sym{***} \\
\midrule

\multicolumn{7}{c}{\textit{Panel D. Shuffle Control} $(n=1000)$} \\
\midrule
$\nu=0.1$ & $0.0112$ & $0.0124$ & $0.0119$ & $0.0401$ & $0.0107$\sym{***} & $0.0108$\sym{***} \\
$\nu=0.2$ & $0.0111$ & $0.0123$ & $0.0118$ & $0.0622$ & $0.0105$\sym{***} & $0.0106$\sym{***} \\
$\nu=0.3$ & $0.0112$ & $0.0124$ & $0.0119$ & $0.0730$ & $0.0106$\sym{***} & $0.0107$\sym{***} \\
$\nu=0.4$ & $0.0112$ & $0.0123$ & $0.0119$ & $0.0783$ & $0.0106$\sym{***} & $0.0107$\sym{***} \\
\bottomrule
\end{tabular}

\caption{OOS MSE for cross-correlation estimators on financial data (2017-2024). In each OOS year, estimators are trained on an expanding sample from 1995 up to the preceding year; results average 1{,}000 runs per test year. Panels A and C use chronological splits, whereas Panels B and D use shuffled dates. Panels A-B vary the number of assets at $\nu=0.25$; Panels C-D vary $\nu$ at $n=1000$. All panels are tested with $\Delta t_\textrm{in} = 500$. Stars identify the estimator, or statistically indistinguishable best group, with the lowest mean loss in each row. Statistical significance is assessed using the paired date-level bootstrap described in \ref{app:bootstrap}, with loss differentials computed against all non-starred alternatives: * $p<0.10$, ** $p<0.05$, and *** $p<0.01$. Abbreviations: NN = Neural Network estimator; MLE = Maximum-Likelihood Estimator; BBP = Benaych-Georges, Bouchaud, and Potters analytical estimator; CV = Monte Carlo Cross-Validated BBP; FF3 and FF5 = Fama--French three- and five-factor covariance estimators \citep{kennethFrenchDataLibrary,fan2008high}.}
\label{tab:MSE_real}
\end{table}

Table \ref{tab:MSE_real} reports the OOS MSE as a function of the total number of assets $n$ (panel A) and the relative dimension $\nu$ (panel C). Across both the $n$- and $\nu$-sweeps, the proposed NN achieves the lowest error. Importantly, the NN remains robust for $n>500$, beyond the training range ($n\le 500$); at $n=1000$ its advantage persists across the full range of relative dimensions $\nu$ (panel C), indicating that the gain is not confined to a specific aspect ratio. For small universes (e.g., $n=250$) it is statistically indistinguishable from BBP, but as $n$ grows BBP deteriorates and can underperform MLE. This behavior is consistent with a macroscopic market mode in real returns, which violates the boundedness conditions underlying BBP. Two diagnostics support this interpretation: the CV estimator does not exhibit the same degradation and remains comparable to the NN, and removing the mode, defined as the daily cross-sectional mean return, largely eliminates the BBP underperformance (\ref{app:ModeRemovalDiagnostic}). When dates are shuffled to suppress temporal non-stationarity, BBP remains unstable while CV continues to track the NN (panels B and D). 

This behavior is consistent with the intended role of the proposed architecture. Under stationarity, the network should act primarily as a spectral denoiser, whereas on financial data it can interpolate between denoising and a learned correction that exploits persistent structure beyond sampling noise. This interpretation is further supported by the canonical-correlation feasibility diagnostic reported in \ref{app:FeasibilityDiagnostic}.

\subsection{Portfolio Replication via Tracking-Error Minimization}\label{sec:tracking_error}
A direct financial use of a rectangular cross-covariance estimator is present in portfolio replication across two distinct universes. Let $\mathbf{w}_Y$ denote a target portfolio on universe $Y$, and let $\mathbf{w}_X$ denote a replicating portfolio on universe $X$, where the two sets of assets are non-overlapping. For fixed $\mathbf{w}_Y$, the relevant objective is to minimize the tracking-error variance $\mathrm{Var}\!\left[\mathbf{w}_X^\top \mathbf{r}_X-\mathbf{w}_Y^\top \mathbf{r}_Y\right]$. Under the budget constraint $\mathbf{1}^\top \mathbf{w}_X=1$, the corresponding closed-form solution is
\begin{equation}
\mathbf{w}_X
=
\boldsymbol{\Sigma}_{XX}^{-1}\boldsymbol{\Sigma}_{XY}\mathbf{w}_Y
-
\boldsymbol{\Sigma}_{XX}^{-1}\mathbf{1}\,
\frac{\mathbf{1}^{\top}\boldsymbol{\Sigma}_{XX}^{-1}\boldsymbol{\Sigma}_{XY}\mathbf{w}_Y-1}
{\mathbf{1}^{\top}\boldsymbol{\Sigma}_{XX}^{-1}\mathbf{1}}.
\end{equation}

The NN formulation allows the proposed cross-correlation module to be embedded in a larger end-to-end architecture. 

In the implementation considered here, RIEnet \parencite{rienet2025,bongiorno2025end} is used for return preprocessing, marginal-volatility estimation, and filtering of $\hat{\boldsymbol{\Sigma}}_{XX}^{-1}$, while the present method provides the cleaned $\hat{\boldsymbol{C}}_{XY}$; the resulting blocks are then used directly in the tracking-error objective. Importantly, non-negativity of $\mathbf{w}_X$ is not enforced during training. At evaluation, all estimators are passed through the same long-only constrained quadratic program. This tests whether the learned estimator remains effective under a stricter implementable rule, rather than adapting to one constraint set. Full implementation details are given in \ref{app:tracking_error}.

In this experiment, the target portfolio on $Y$ is Equally Weighted (EW), and we use a yearly expanding-window scheme in which each model is trained up to the previous year-end and tested on the following calendar year. To reduce survivorship bias, we use an OOS horizon of $\Delta t_{\mathrm{out}}=10$ daily observations. The evaluation universe contains $n=500$ stocks with $\nu:=n_X/(n_X+n_Y)=0.40$, corresponding to a 200-stock replication universe $X$ and a 300-stock target universe $Y$, with $\Delta t_{\mathrm{in}}=500$ in-sample daily observations. We compare the proposed NN estimator with the EW benchmark, Fama--French three- and five-factor benchmarks \citep{kennethFrenchDataLibrary,fan2008high}, the sample estimator, and two hybrid procedures combining Quadratic Inverse Shrinkage (QIS) cleaning~\parencite{ledoit2022quadratic} of $\boldsymbol{\Sigma}_{XX}$ with CV or BBP cleaning of $\boldsymbol{\Sigma}_{XY}$.

Table~\ref{tab:tracking_error} reports the tracking-error results separately for each annual test period. The NN estimator achieves the lowest OOS tracking error in every year from 2017 to 2024, with its advantage particularly pronounced during the 2020 COVID-19 market-stress period. Averaged across years, it reduces tracking error by approximately $10\%$ relative to the best non-neural benchmark. This ranking is somewhat counterintuitive: the hybrid estimators built from MSE-oriented cleaning components do not improve on the sample benchmark in this downstream task and can even underperform the EW portfolio. Thus, improving an auxiliary matrix loss is not sufficient for portfolio replication. By contrast, once the model is trained on a closely aligned analytical tracking-error objective, the resulting filtered blocks remain effective even after they are mapped into a stricter long-only constrained portfolio problem. This transfer is useful because in practice the final constraint set is mandate-dependent and may require numerical procedures with no simple closed-form solution. A useful estimator should therefore remain valid beyond the specific constraint set used during training.

\begin{table}[t]
\centering
\small

\centering
\providecommand{\sym}[1]{\textsuperscript{#1}}
\footnotesize
\setlength{\tabcolsep}{2.5pt}
\renewcommand{\arraystretch}{0.93}

\begin{tabular}{@{}llllllll@{}}
\toprule
Year & EW & MLE & FF3 & FF5 & CV & BBP & NN \\
\midrule
\multicolumn{8}{c}{\textit{Year-by-Year OOS Tracking Error (2017--2024)}} \\
\midrule
2017 & $0.01961$ & $0.01965$ & $0.02251$ & $0.02152$ & $0.02608$ & $0.02855$ & $0.01757$\sym{***} \\
2018 & $0.02202$ & $0.02206$ & $0.02462$ & $0.02208$ & $0.02886$ & $0.02878$ & $0.01981$\sym{***} \\
2019 & $0.02167$ & $0.02152$ & $0.02178$ & $0.02046$ & $0.03016$ & $0.03127$ & $0.01929$\sym{***} \\
2020 & $0.03326$ & $0.03171$ & $0.03265$ & $0.03194$ & $0.04852$ & $0.09355$ & $0.02867$\sym{***} \\
2021 & $0.02636$ & $0.02490$ & $0.02633$ & $0.02585$ & $0.03805$ & $0.06448$ & $0.02257$\sym{***} \\
2022 & $0.02798$ & $0.02577$ & $0.02848$ & $0.02640$ & $0.03875$ & $0.05484$ & $0.02342$\sym{***} \\
2023 & $0.02425$ & $0.02381$ & $0.02456$ & $0.02323$ & $0.03346$ & $0.04557$ & $0.02146$\sym{***} \\
2024 & $0.02140$ & $0.02002$ & $0.02175$ & $0.01927$ & $0.02524$ & $0.04953$ & $0.01765$\sym{***} \\
\midrule
Mean & $0.02457$ & $0.02368$ & $0.02534$ & $0.02384$ & $0.03364$ & $0.04957$ & $0.02130$\sym{***} \\
\bottomrule
\end{tabular}

\caption{Year-by-year OOS TE results over the 8 annual test periods from 2017 to 2024. Each annual estimate is based on 3{,}200 sampled tracking instances. Stars identify the estimator with the lowest tracking error in each row. Statistical significance is assessed using the paired date-level bootstrap described in \ref{app:bootstrap}, with loss differentials computed against all non-starred alternatives: * $p<0.10$, ** $p<0.05$, and *** $p<0.01$. Abbreviations: NN = Neural Network estimator; MLE = Maximum-Likelihood Estimator; EW = Equally Weighted; FF3 and FF5 = Fama-French three- and five-factor covariance estimator \citep{kennethFrenchDataLibrary,fan2008high}; BBP = Benaych-Georges, Bouchaud, and Potters analytical estimator; CV = Monte Carlo Cross-Validated BBP.}\label{tab:tracking_error}

\end{table}

\section{Discussion and Conclusion}\label{sec:disc_conc} 
The results separate stationary denoising from prediction under changing dependence. On the synthetic benchmarks and shuffled returns, the estimator behaves as an RMT-consistent denoiser and remains close to BBP or the cross-validated benchmark. On chronological equity data, it remains stable as the universe grows, while analytical BBP shrinkage deteriorates in the presence of a macroscopic market mode. In portfolio replication, it also attains the lowest annual OOS tracking error in every test year from 2017 to 2024. These findings indicate that the structured neural map preserves the analytical behavior in favorable regimes and adds a useful correction when dependence is time-varying.

We validate the method on synchronous cross-covariances and portfolio replication. The same rectangular object may also support cross-asset signals or alpha transfer when information is measured on one block and implemented through another. Future work can extend the symmetry-preserving construction to lead-lag matrices and train it with decision-aware objectives matched to the final trading problem.

\section*{Acknowledgments}
This work was performed using HPC resources from the ``Mésocentre'' computing center of CentraleSupélec and École Normale Supérieure Paris-Saclay supported by CNRS and Région Île-de-France (\url{http://mesocentre.centralesupelec.fr/}).
EM~acknowledges a fellowship  funded  by PNRR  for the PhD  DOT1608375  in Sistemi complessi per le scienze fisiche, socio-economiche e della vita of Catania University.  RNM acknowledge financial support of the Italian PRIN research project 2017WZFTZP “Stochastic forecasting in complex systems” funded by the Italian Ministero dell'Università e della Ricerca.

\section*{Declaration of Generative AI and AI-assisted technologies in the writing process}
During the preparation of this work the authors used Grammarly in order to improve language and readability. After using this tool/service, the authors reviewed and edited the content as needed and take full responsibility for the content of the published article.

  \printbibliography[title={References}]
\end{refsection}

\begin{refsection}
\appendix

\section{BBP Analytical Cleaner as a Limiting Case of the Two-Stream NN}
\label{app:BBP_as_limit}

This section rewrites the analytical BBP singular-value cleaner of \textcite{benaych2019optimal} using the notation of Section~\ref{sec:architecture}, and clarifies why the proposed two-stream neural architecture contains the analytical mapping as a special case.

\subsection{BBP Analytical Shrinkage: Algorithm~1}
\label{app:BBP_algo1}
Let $\{\mathbf{x}(t)\in\mathbb{R}^{n_x},\mathbf{y}(t)\in\mathbb{R}^{n_y}\}_{t=1}^{\Delta t_\textrm{in}}$ denote standardized observations, and let $\widehat{\mathbf{C}}_{XY}$, $\widehat{\mathbf{C}}_{XX}$, and $\widehat{\mathbf{C}}_{YY}$ be the corresponding empirical cross-correlation and marginal correlation matrices. We write
\begin{equation}
\widehat{\mathbf{C}}_{XY}=\sum_{k=1}^{r}\widehat{s}_k\,\widehat{\mathbf{u}}_k\widehat{\mathbf{v}}_k^\top,\qquad r:=\min(n_x,n_y),
\end{equation}
and define the marginal projections
\begin{equation}
\widehat{\gamma}^{(x)}_k:=\widehat{\mathbf{u}}_k^\top \widehat{\mathbf{C}}_{XX}\widehat{\mathbf{u}}_k,
\qquad
\widehat{\gamma}^{(y)}_k:=\widehat{\mathbf{v}}_k^\top \widehat{\mathbf{C}}_{YY}\widehat{\mathbf{v}}_k.
\end{equation}
When $n_x\neq n_y$, BBP also uses the aggregate marginal energy in the extra directions of the larger side. Assuming $n_x\le n_y$ without loss of generality, we set
\begin{equation}
\widehat{\gamma}^{(y)}_{r+}:=\sum_{\ell=r+1}^{n_y}\widehat{\mathbf{v}}_\ell^\top \widehat{\mathbf{C}}_{YY}\widehat{\mathbf{v}}_\ell,
\end{equation}
where $\{\widehat{\mathbf{v}}_\ell\}_{\ell=r+1}^{n_y}$ completes the empirical right singular vectors to an orthonormal basis. For each $k$, define $\zeta_k:=\widehat{s}_k+i\eta$ with $\eta:=(r n_y\Delta t_\textrm{in})^{-1/12}$ and
\begin{align}
H(\zeta_k) &:= \frac{1}{\Delta t_\textrm{in}}\sum_{\ell=1}^{r}\frac{\widehat{s}_\ell^2}{\zeta_k^{2}-\widehat{s}_\ell^{2}}, \\
A(\zeta_k) &:= \frac{1}{\Delta t_\textrm{in}}\sum_{\ell=1}^{r}\frac{\widehat{\gamma}^{(x)}_\ell}{\zeta_k^{2}-\widehat{s}_\ell^{2}}, \\
B(\zeta_k) &:= \frac{1}{\Delta t_\textrm{in}}\left(\sum_{\ell=1}^{r}\frac{\widehat{\gamma}^{(y)}_\ell}{\zeta_k^{2}-\widehat{s}_\ell^{2}}+\zeta_k^{-2}\widehat{\gamma}^{(y)}_{r+}\right).
\end{align}
With
\begin{equation}
\Theta(\zeta_k):=\frac{\zeta_k^2 A(\zeta_k)B(\zeta_k)}{1+H(\zeta_k)},\qquad
L(\zeta_k):=H(\zeta_k)-\frac{\Theta(\zeta_k)}{1+H(\zeta_k)-\Theta(\zeta_k)},
\end{equation}
the analytical cleaned singular values are
\begin{equation}
\widehat{s}^{\mathrm{BBP}}_k:=\widehat{s}_k\left(\frac{\textrm{Im}L(\zeta_k)}{\textrm{Im}H(\zeta_k)}\right)_+,
\qquad (x)_+:=\max\{x,0\}.
\label{eq:BBP_shrinkage}
\end{equation}
The BBP rotationally invariant estimator is then reconstructed as $\widehat{\mathbf{C}}^{\mathrm{BBP}}_{XY}:=\sum_{k=1}^{r}\widehat{s}^{\mathrm{BBP}}_k\widehat{\mathbf{u}}_k\widehat{\mathbf{v}}_k^\top$.

\subsection{Relationship Between the Two-Stream NN and the Analytical Shrinkage}
\label{app:BBP_in_NN}
The BBP map in Eq.~\eqref{eq:BBP_shrinkage} depends on the observations only through the empirical singular spectrum, the marginal projections, and the aspect ratios $q_x=n_x/\Delta t_\textrm{in}$ and $q_y=n_y/\Delta t_\textrm{in}$. Once the empirical singular vectors are fixed, BBP is therefore a deterministic permutation-equivariant mapping from $(\widehat{\mathbf{s}},\widehat{\boldsymbol{\gamma}}^{(x)},\widehat{\boldsymbol{\gamma}}^{(y)},q_x,q_y)$ to the cleaned singular values.

The NN is constructed precisely on these sufficient statistics. The shared token-wise encoder can pass the local quantities $(s_k,\gamma^{(x)}_k,\gamma^{(y)}_k,q_x,q_y)$ into the latent representation. The bidirectional LSTM provides global spectral context, which is required by the sums defining $H$, $A$, and $B$. Finally, the pointwise head can approximate the continuous scalar algebra mapping this state to the residual
\begin{equation}
\delta_k^{\mathrm{BBP}}:=\widehat{s}^{\mathrm{BBP}}_k-\widehat{s}_k.
\end{equation}
Thus, there exists a parameter configuration $\theta_{\mathrm{BBP}}$ for which
\begin{equation}
\widetilde{s}_k(\theta_{\mathrm{BBP}})=\widehat{s}^{\mathrm{BBP}}_k,\qquad k=1,\ldots,r,
\end{equation}
on the stationary analytical domain where BBP is defined. The architecture is therefore a symmetry-preserving hypothesis class for rotationally invariant shrinkage: in stationary regimes it can specialize to analytical denoising, while under non-stationarity it can deviate from that baseline through learned corrective terms.

\subsection{Admissible Singular Values Under Fixed Marginals}
\label{app:FeasibleSV}
A candidate cross-correlation block $\mathbf{C}_{XY}\in\mathbb{R}^{n_x\times n_y}$ is feasible under fixed marginal correlation matrices $\mathbf{C}_{XX}$ and $\mathbf{C}_{YY}$ if the block matrix
\begin{equation}
\mathbf{C}:=\begin{pmatrix}\mathbf{C}_{XX}&\mathbf{C}_{XY}\\ \mathbf{C}_{XY}^\top&\mathbf{C}_{YY}\end{pmatrix}
\label{eq:block_corr}
\end{equation}
is positive semidefinite \parencite{muirhead2009aspects}. If $\mathbf{C}_{XX}\succ0$ and $\mathbf{C}_{YY}\succ0$, define the whitened block
\begin{equation}
\mathbf{C}^{(w)}_{XY}:=\mathbf{C}_{XX}^{-1/2}\mathbf{C}_{XY}\mathbf{C}_{YY}^{-1/2}.
\label{eq:whitened_def}
\end{equation}
A congruence transformation reduces feasibility to the identity-marginal block matrix. By the Schur complement \parencite{ouellette1981schur}, this is equivalent to
\begin{equation}
0\le s_k^{(w)}\le 1,\qquad k=1,\ldots,r,
\label{eq:sv_bound_whitened}
\end{equation}
where $s_k^{(w)}$ are the singular values of $\mathbf{C}^{(w)}_{XY}$. The strict inequality $s_k^{(w)}<1$ is sufficient for positive definiteness. This feasibility condition is distinct from a pure shrinkage condition $\widetilde{s}_k\le \widehat{s}_k$ in the unwhitened spectrum. The latter is natural for stationary denoising, whereas an OOS forecasting objective may require amplifying selected empirical modes while still satisfying the canonical-correlation bound in Eq.~\eqref{eq:sv_bound_whitened}. This distinction is evaluated empirically in \ref{app:FeasibilityDiagnostic}.

\section{Cross-validated Singular-Value Shrinkage}
\label{app:CV}
\subsection{Boundedness in the Presence of a Mode}
\label{app:mode_problem}
The analytical BBP shrinkage is derived under high-dimensional assumptions that include boundedness of the operator norm of the population covariance of the concatenated vector. This condition can fail in the presence of a strong common mode. For instance, the equicorrelation matrix $\mathbf{A}_n=(1-m)\mathbf{I}_n+m\mathbf{1}\mathbf{1}^\top$, with $0<m<1$, has leading eigenvalue $1+m(n-1)$ and therefore $\|\mathbf{A}_n\|_{\mathrm{op}}\sim mn$. If such a component is embedded in the population covariance, the bounded-spectrum assumption underlying the analytical derivation is violated. We therefore introduce a cross-validated (CV) singular-value cleaner that provides a stationary BBP-like benchmark without relying on this boundedness condition.

\subsection{Cross-Validation Algorithm}
Let $\{\mathbf{x}_t,\mathbf{y}_t\}_{t=1}^{\Delta t_\textrm{in}}$ denote standardized paired observations. We form
\begin{equation}
\widehat{\mathbf{C}}_{XY}=\frac{1}{\Delta t_\textrm{in}}\sum_{t=1}^{\Delta t_\textrm{in}}\mathbf{x}_t\mathbf{y}_t^\top=\widehat{\mathbf{U}}\,\operatorname{Diag}(\hat{\mathbf{s}})\widehat{\mathbf{V}}^\top.
\end{equation}
Within the rotationally invariant class, the oracle diagonal is $s_k^\star=\widehat{\mathbf{u}}_k^\top\mathbf{C}_{XY}\widehat{\mathbf{v}}_k$. CV estimates this diagonal by splitting the time indices into $\kappa$ folds. For fold $f$, the singular directions are estimated from $\widehat{\mathbf{C}}^{(f)}_{XY,\mathrm{train}}$ and evaluated on $\widehat{\mathbf{C}}^{(f)}_{XY,\mathrm{test}}$ through
\begin{equation}
\tilde{s}^{(f)}_k:=\widehat{\mathbf{u}}_{\mathrm{train},k}^{(f)\top}\widehat{\mathbf{C}}^{(f)}_{XY,\mathrm{test}}\widehat{\mathbf{v}}_{\mathrm{train},k}^{(f)},
\qquad
\tilde{s}^{\mathrm{CV}}_k:=\frac{1}{\kappa}\sum_{f=1}^{\kappa}\tilde{s}^{(f)}_k.
\end{equation}
We optionally enforce monotonicity by isotonic regression and reconstruct $\widetilde{\mathbf{C}}^{\mathrm{CV}}_{XY}$ in the empirical singular-vector basis of the full in-sample matrix. This procedure is a stationary holdout estimator of mode strengths, related in spirit to covariance holdout and NERCOME-type regularization \parencite{AbadirDistasoZikes2014,Lam2016NERCOME}.

\section{Synthetic Data Validation}
\label{app:Synthetic}
Most benchmarks are based on \textcite{benaych2019optimal}. They test the proposed NN shrinkage estimator across finite-rank, heavy-tailed, and mode-dominated spectral regimes. We additionally include a finance-style conditional multifactor DCC-GARCH benchmark.

\subsection{Synthetic Benchmarks}
\subsubsection*{Benchmark I: Finite-Rank Spiked Models}
The first benchmark corresponds to Models 1--5 in \textcite{benaych2019optimal}. We consider a Gaussian block model with population covariance
\begin{equation}
\boldsymbol{\Sigma}:=\begin{pmatrix}\mathbf{I}_{n_x}&\boldsymbol{\Sigma}_{XY}\\ \boldsymbol{\Sigma}_{XY}^\top&\boldsymbol{\Sigma}_{XY}^\top\boldsymbol{\Sigma}_{XY}+\sigma^2\mathbf{I}_{n_y}\end{pmatrix},
\end{equation}
where $\sigma^2=0.5$. The cross-block $\boldsymbol{\Sigma}_{XY}$ has Haar-distributed singular vectors and a fraction $\xi$ of nonzero singular values, drawn independently from $\mathrm{Unif}[0.2,0.5]$. Panel~A of Table~\ref{tab:synthetic_benchmarks} reports performance as $\xi$ varies.

\subsubsection*{Benchmark II: Heavy-Tailed Bulk Models}
\label{appsec:Bench2}
The second benchmark corresponds to Models 6--10 in \textcite{benaych2019optimal}. We draw a rectangular matrix $\mathbf{W}\in\mathbb{R}^{n\times 2n}$, with $n:=n_x+n_y$, either from a Gaussian distribution or from a symmetric heavy-tailed law with tail exponent $\alpha$, and set
\begin{equation}
\boldsymbol{\Sigma}:=\frac{1}{2n}\mathbf{W}\mathbf{W}^\top.
\end{equation}
The blocks $\boldsymbol{\Sigma}_{XX}$, $\boldsymbol{\Sigma}_{XY}$, and $\boldsymbol{\Sigma}_{YY}$ are extracted under the $X/Y$ partition. Lower $\alpha$ corresponds to heavier tails and a more dispersed spectrum. Panel~B reports the resulting estimator comparison.

\subsubsection*{Benchmark III: White Heavy-Tailed Cross-Correlation Models}
\label{appsec:Bench3}
This benchmark isolates the cross-block estimation problem by retaining the cross-block from Benchmark~II but replacing the two marginal blocks by scaled identities:
\begin{equation}
\boldsymbol{\Sigma}:=\begin{pmatrix}s_{\max}\mathbf{I}_{n_x}&\boldsymbol{\Sigma}_{XY}\\ \boldsymbol{\Sigma}_{XY}^\top&s_{\max}\mathbf{I}_{n_y}\end{pmatrix},
\qquad s_{\max}:=\|\boldsymbol{\Sigma}_{XY}\|_2.
\end{equation}
This removes marginal anisotropy while preserving the inherited heavy-tailed cross-block structure. Panel~C reports results as $\alpha$ varies.

\subsubsection*{Benchmark IV: Gaussian Bulk Models with Common Mode}
\label{appsec:Bench4}
This benchmark starts from the Gaussian instance of Benchmark~II, with $\boldsymbol{\Sigma}^{(G)}=(2n)^{-1}\mathbf{W}\mathbf{W}^\top$. To emulate a positive market mode, we apply the congruence transformation
\begin{equation}
\mathbf{M}:=(1-m)\mathbf{I}_n+m\mathbf{1}\mathbf{1}^\top,
\qquad
\boldsymbol{\Sigma}^{(M)}=\mathbf{M}^{1/2}\boldsymbol{\Sigma}^{(G)}\mathbf{M}^{1/2},
\end{equation}
and then normalize to correlation form with $\mathbf{D}:=\operatorname{Diag}(\operatorname{diag}(\boldsymbol{\Sigma}^{(M)}))$ and $\boldsymbol{\Sigma}=\mathbf{D}^{-1/2}\boldsymbol{\Sigma}^{(M)}\mathbf{D}^{-1/2}$. Panel~D reports performance as $m$ varies.

\subsubsection*{Benchmark V: Conditional FF-Style Factor-DCC-GARCH Models}
\label{app:FactorDCC}
The fifth benchmark adds a Fama--French-style conditional multifactor data-generating process. The factors are simulated, and the labels FF-3, FF-4, and FF-5 indicate only the number and role of the synthetic factors. Let $n:=n_x+n_y$ and let $d_f\in\{3,4,5\}$ denote the number of latent factors. Returns are generated as
\begin{equation}
\mathbf{r}_t=\mathbf{A}\mathbf{f}_t+\boldsymbol{\varepsilon}_t,
\qquad
\mathbf{A}=\mathbf{U}\boldsymbol{\Lambda}^{1/2},
\qquad
\mathbf{U}^\top\mathbf{U}=\mathbf{I}_{d_f},
\end{equation}
where $\boldsymbol{\Lambda}:=\operatorname{Diag}(\lambda_1,\ldots,\lambda_{d_f})$. The first loading direction is constrained to be positive and represents a market mode, whereas the remaining $d_f-1$ directions are sign-changing long-short style directions. Thus, FF-3 contains one market direction and two style directions, FF-4 adds one additional style direction in the spirit of \textcite{carhart1997persistence}, and FF-5 follows the five-factor terminology of \textcite{fama2015five}. No historical Kenneth French factor returns are used.

The factor covariance is conditionally time-varying. We write
\begin{equation}
\mathbf{f}_t\mid\mathcal{F}_{t-1}\sim(\mathbf{0},\mathbf{H}_{f,t}),
\qquad
\mathbf{H}_{f,t}=\mathbf{D}_{f,t}\mathbf{R}_{f,t}\mathbf{D}_{f,t},
\end{equation}
with
\begin{equation}
\mathbf{D}_{f,t}:=\operatorname{Diag}(\sqrt{h_{1,t}},\ldots,\sqrt{h_{d_f,t}}).
\end{equation}
The factor variances follow GARCH recursions \parencite{engle1982autoregressive,bollerslev1986generalized},
\begin{equation}
h_{j,t}=\omega_j+\alpha_f f_{j,t-1}^2+b_f h_{j,t-1},
\qquad
\alpha_f+b_f<1,
\qquad j=1,\ldots,d_f,
\end{equation}
and the factor correlation matrix follows the DCC recursion of \textcite{engle2002dynamic},
\begin{equation}
\mathbf{Q}_{f,t}=(1-a-b)\bar{\mathbf{Q}}_f+a\mathbf{z}_{t-1}\mathbf{z}_{t-1}^\top+b\mathbf{Q}_{f,t-1},
\end{equation}
\begin{equation}
\mathbf{q}_{f,t}:=\operatorname{diag}(\mathbf{Q}_{f,t}),
\qquad
\mathbf{R}_{f,t}=\operatorname{Diag}(\mathbf{q}_{f,t})^{-1/2}\mathbf{Q}_{f,t}\operatorname{Diag}(\mathbf{q}_{f,t})^{-1/2},
\end{equation}
where $a+b<1$ and $\mathbf{z}_t=\mathbf{D}_{f,t}^{-1}\mathbf{f}_t$. This construction follows ARCH/GARCH and DCC covariance modeling and is consistent with factor-ARCH formulations \parencite{diebold1989factor,engle1990asset}. It introduces persistence through conditional second moments rather than through material return autocorrelation.

Let $\mathbf{A}_X$ and $\mathbf{A}_Y$ denote the rows of $\mathbf{A}$ associated with the two synchronous blocks. Since idiosyncratic shocks are independent across assets, the conditional population cross-covariance block is
\begin{equation}
\boldsymbol{\Sigma}_{XY,t}=\mathbf{A}_X\mathbf{H}_{f,t}\mathbf{A}_Y^\top.
\end{equation}
The corresponding population cross-correlation block is obtained by rescaling $\boldsymbol{\Sigma}_{XY,t}$ by the conditional marginal standard deviations. In the reported experiment, the supervised target is the average population cross-correlation block over the subsequent OOS window.

\subsection{Training on Synthetic Data}
\label{sec:synthetic_training}
We train a separate NN model for each synthetic benchmark. Training batches are generated on the fly, which exposes the model to a non-repeated stream of simulated observations and reduces memorization of a finite training set. The calibration ranges are chosen so that selected test parameters are outside the training distribution. For Panels A--D, $n\sim\mathcal{U}[50,500]$, $\Delta t_\textrm{in}\sim\mathcal{U}[200,1200]$, and $\nu\sim\mathcal{U}[0.05,0.95]$; the benchmark-specific parameters are $\xi\sim\mathcal{U}[0.2,0.35]$ for Benchmark I, $\alpha\sim\mathcal{U}[1.5,2.5]$ for Benchmarks II--III, and $m\sim\mathcal{U}[0.2,0.3]$ for Benchmark IV. All synthetic models use Adam with initial learning rate $0.001$, exponential decay $0.99$ per epoch, 50 epochs, base batch size 32, and gradient accumulation over four steps.

For Benchmark V, we train one NN estimator for each value of $d_f\in\{3,4,5\}$. Training batches randomize $n\in[50,500]$, $\Delta t_{\mathrm{in}}\in[200,1200]$, and $\nu=n_x/(n_x+n_y)\in[0.05,0.95]$. Each training sample is generated from an independent factor-DCC path after burn-in, and the model receives raw return matrices from which the empirical marginal correlations and cross-correlation are computed internally. The implementable estimators are MLE, BBP, CV, and NN. The simulation also computes a DCC population oracle, which has zero error by construction under the population-target design and is therefore omitted from Table~\ref{tab:synthetic_benchmarks}. Following the paired loss-comparison logic of \textcite{diebold1995comparing}, the NN has the lowest mean MSE for all three multifactor conditions, and the paired comparisons against MLE, BBP, and CV reject equal average loss in each case.

For testing, Table~\ref{tab:synthetic_benchmarks} uses $n_x=200$, $n_y=350$, and $\Delta t_\textrm{in}=500$ as in \textcite{benaych2019optimal}. Panel E uses $n=550$ and is therefore slightly outside the training dimension range, with $\Delta t_{\mathrm{out}}=240$ and $1{,}000$ Monte Carlo samples for each factor specification.

\section{Market Data Validation}
\label{app:real}
\subsection{Universe Selection}
\label{sec:Data}
The data filtering procedure follows \textcite{bongiorno2025end}. The dataset comprises NYSE- and NASDAQ-listed equities from January 1, 1990 to December 31, 2024, including ADRs, and excludes ETFs, funds, and non-operating vehicles. Only common shares are retained. The eligible universe on each day is determined using information available strictly prior to that day, and securities that will be delisted within the subsequent $\Delta t_{\mathrm{out}}=240$ trading days are excluded in the MSE experiment. Each stock must have a stable five-year trading history; within every rolling one-year subwindow, it must participate in the closing auction on at least $95\%$ of trading days. Additional liquidity filters require closing-auction execution, sufficient traded volume relative to shares outstanding and market capitalization, price between 10 and 2000 USD, and at least 5 million shares outstanding. We also remove near-zero variance stocks, retain only one share class per issuer, and remove the lower-capitalization member of pairs with in-sample correlation above $0.95$. After filtering, the top 1000 stocks by market capitalization are retained each day. The final panel spans 1995--2024, after the five-year lookback requirement.

\subsubsection{Delisting Risk and Alternative Remedies}
\label{app:delisting}
The MSE experiment restricts the OOS universe to names observed throughout the one-year OOS target window, which avoids mechanically missing observations but introduces mild survivorship conditioning. One remedy is to shorten the OOS horizon to a window that does not exceed the public-notice period preceding exchange delisting. Delisting through Form 25 becomes effective 10 days after filing and exchanges must provide public notice no fewer than 10 days in advance \parencite{sec12d22}; we adopt this shorter horizon in the tracking-error application. A second remedy is to compute pairwise correlations over overlapping observations, which avoids full-horizon survivorship conditioning but can yield indefinite correlation estimates. A third possibility is to quantify sensitivity to progressively less restrictive OOS availability requirements. We do not pursue the latter extensions here because the present objective is estimator comparison rather than a fully specified trading backtest.

\subsection{Training on Real Market Data}
\label{app:Training}
The model is trained under a walk-forward protocol on U.S. equities, with OOS evaluation spanning 2017--2024. At each iteration, the network is trained from scratch on data from 1995 up to the year immediately preceding the corresponding test segment. In each training batch, the number of assets is sampled as $n\sim\mathcal{U}[50,500]$, the split fraction as $\nu\sim\mathcal{U}[0.05,0.95]$, and the lookback length as $\Delta t_\textrm{in}\sim\mathcal{U}[200,1200]$. The inputs are $\widehat{\mathbf{C}}_{XX}$, $\widehat{\mathbf{C}}_{YY}$, and $\widehat{\mathbf{C}}_{XY}$ computed from daily adjusted-close returns. The model outputs a cleaned estimate of $\mathbf{C}_{XY}$ and is trained by minimizing MSE against the realized OOS target $\mathbf{C}^{\mathrm{OOS}}_{XY}$ computed over $\Delta t_{\mathrm{out}}=240$ trading days.

For shuffled controls, we merge the in-sample and OOS segments into a pool of trading days, randomly permute the dates, and re-split into in-sample and OOS windows of the same lengths. This destroys chronological structure while preserving the empirical daily cross-sectional distribution, so IS and OOS blocks differ only through finite-sample noise. Optimization uses Adam with initial learning rate $0.001$, exponential decay $0.99$, 50 epochs, base batch size 32, and four-step gradient accumulation.

\subsection{Market-Mode Removal as a Boundedness Diagnostic}
\label{app:ModeRemovalDiagnostic}
\begin{table}[tbh]
\centering
\centering
\providecommand{\sym}[1]{\textsuperscript{#1}}
\small
\setlength{\tabcolsep}{2.6pt}
\renewcommand{\arraystretch}{0.95}

\begin{tabular}{@{}lllllll@{}}
\toprule
Condition & MLE & FF3 & FF5 & BBP & CV & NN \\
\midrule
\multicolumn{7}{c}{\textit{Panel B. Shuffle Control after Market-Mode Removal} $(\nu=0.25)$} \\
\midrule
$n=250$  & $0.0110$ & $0.0127$ & $0.0119$ & $0.0100$ & $0.0098$\sym{***} & $0.0099$ \\
$n=500$  & $0.0110$ & $0.0127$ & $0.0118$ & $0.0100$ & $0.0098$\sym{***} & $0.0099$ \\
$n=750$  & $0.0110$ & $0.0127$ & $0.0118$ & $0.0100$ & $0.0097$\sym{***} & $0.0099$ \\
$n=1000$ & $0.0110$ & $0.0127$ & $0.0118$ & $0.0101$ & $0.0097$\sym{***} & $0.0099$ \\
\midrule
\multicolumn{7}{c}{\textit{Panel D. Shuffle Control after Market-Mode Removal} $(n=1000)$} \\
\midrule
$\nu=0.1$ & $0.0110$ & $0.0127$ & $0.0118$ & $0.0100$ & $0.0098$\sym{***} & $0.0100$ \\
$\nu=0.2$ & $0.0110$ & $0.0127$ & $0.0118$ & $0.0100$ & $0.0097$\sym{***} & $0.0099$ \\
$\nu=0.3$ & $0.0110$ & $0.0127$ & $0.0118$ & $0.0102$ & $0.0097$\sym{***} & $0.0099$ \\
$\nu=0.4$ & $0.0110$ & $0.0127$ & $0.0118$ & $0.0102$ & $0.0097$\sym{***} & $0.0099$ \\
\bottomrule
\end{tabular}
\caption{OOS MSE for cross-correlation estimators on financial data (2017--2024) after removing the market mode. For each OOS year, estimators are trained on an expanding sample from 1995 up to the preceding year; reported values average 1{,}000 runs per test year. The table reports the shuffled-date controls corresponding to the $n$-sweep and the $\nu$-sweep: Panel B varies the number of assets at fixed $\nu=0.25$, whereas Panel D varies $\nu$ at fixed $n=1000$. FF3 and FF5 denote Fama--French three- and five-factor covariance estimators \citep{kennethFrenchDataLibrary,fan2008high} evaluated on the same OOS sample. Stars identify the estimator, or statistically indistinguishable best group, with the lowest mean loss in each row. Statistical significance is assessed using the paired date-level bootstrap described in \ref{app:bootstrap}, with loss differentials computed against all non-starred alternatives: * $p<0.10$, ** $p<0.05$, and *** $p<0.01$. Abbreviations: NN = Neural Network estimator; MLE = Maximum-Likelihood Estimator; BBP = Benaych-Georges, Bouchaud, and Potters analytical estimator; CV = Monte Carlo Cross-Validated BBP.}
\label{tab:MSE_no_mode}
\end{table}

To test whether BBP deterioration is driven by a macroscopic market mode, we repeat the real-data pipeline after subtracting the daily cross-sectional mean return. For asset $i$ on day $t$, we define
\begin{equation}
\hat{\mu}_t:=\frac{1}{n}\sum_{i=1}^{n}R_{i,t},\qquad R^{(0)}_{i,t}:=R_{i,t}-\hat{\mu}_t.
\end{equation}
The transformation is applied separately in each in-sample and OOS window. Table~\ref{tab:MSE_no_mode} shows that the sharp BBP deterioration observed in raw returns is substantially reduced once this mode is removed, supporting the interpretation that the leading common component violates the boundedness condition discussed in \ref{app:mode_problem}. The chronological NN advantage is not fully eliminated, indicating that market-mode boundedness and time variation are related but distinct sources of difficulty.

\subsection{Feasibility Diagnostic via Canonical-Correlation Boundedness}
\label{app:FeasibilityDiagnostic}
\begin{figure}[thb]
\centering
\includegraphics[width=0.4\linewidth]{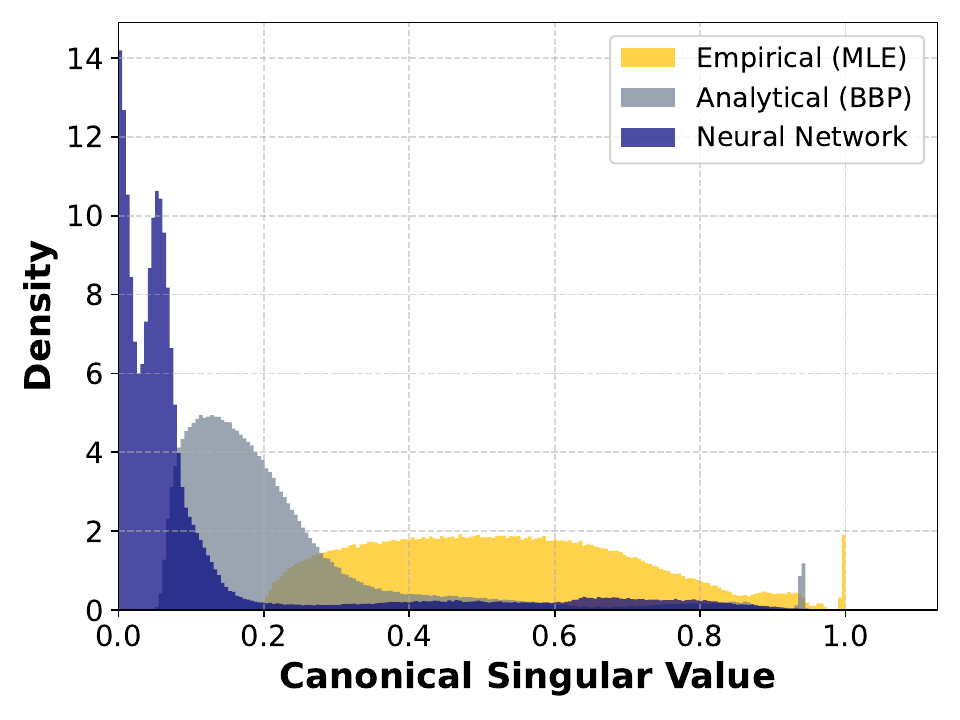}
\includegraphics[width=0.4\linewidth]{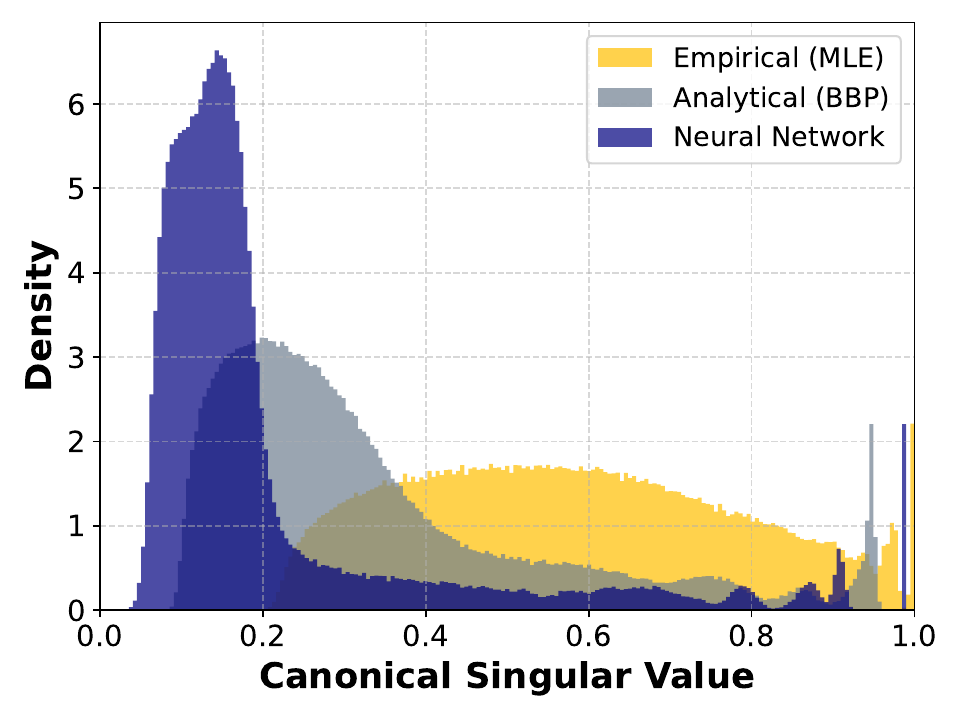}
\includegraphics[width=0.4\linewidth]{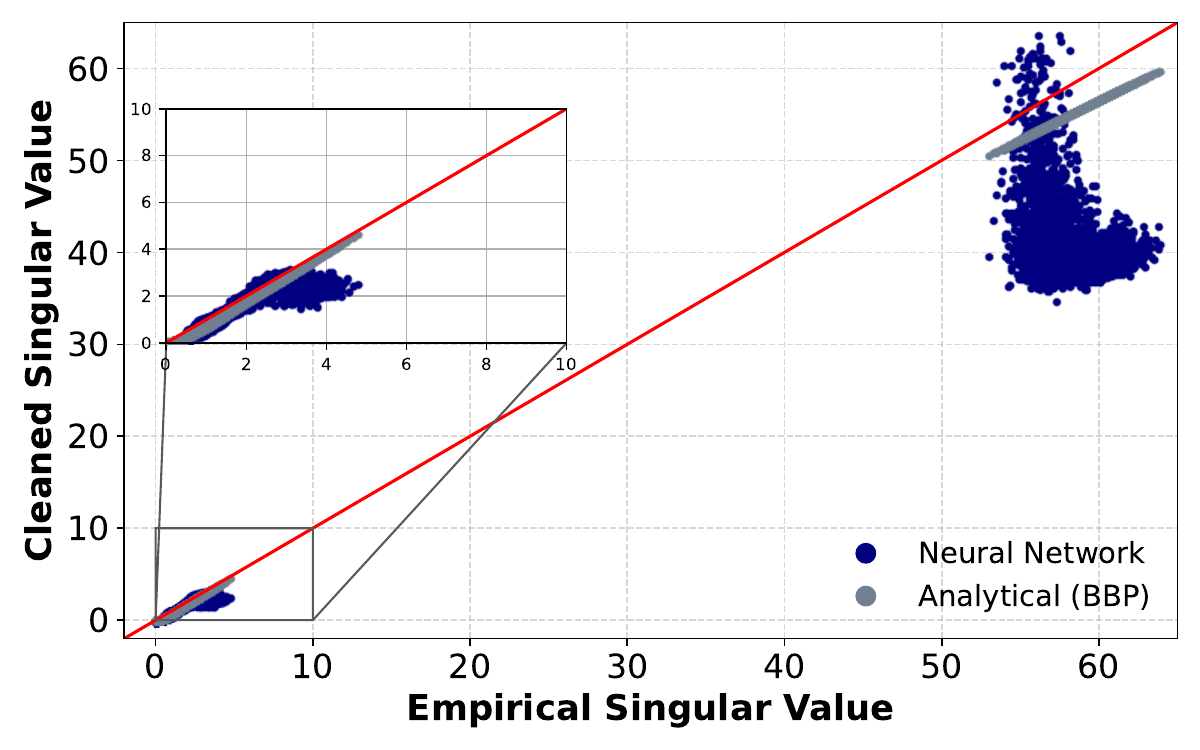}
\includegraphics[width=0.4\linewidth]{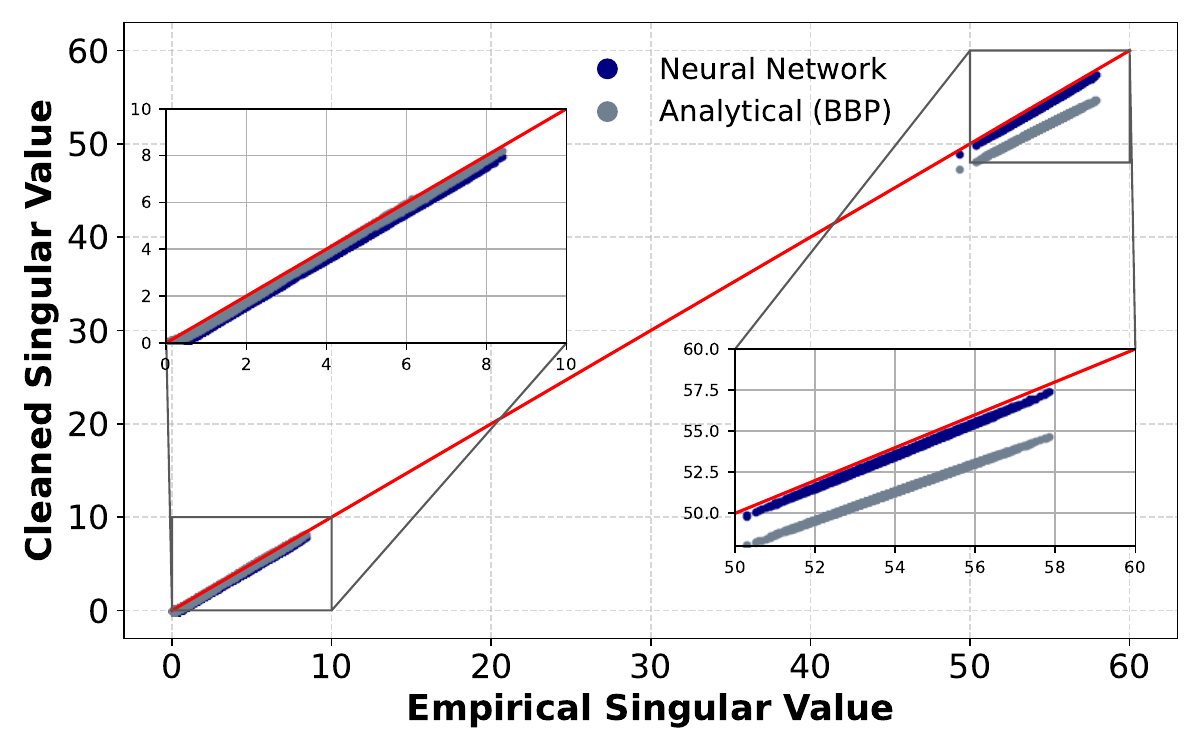}
\caption{Feasibility diagnostic for the reconstructed cross-correlation block. Top row: empirical distributions of canonical singular values of the whitened block $\widetilde{\mathbf{C}}^{(w)}_{XY}$. Bottom row: cleaned singular values in the unwhitened domain plotted against empirical singular values. Left panels use the chronological pipeline; right panels use the shuffled control.}
\label{fig:CCC_boundedness}
\end{figure}

Using the feasibility result in \ref{app:FeasibleSV}, we evaluate whether the neural estimator produces cross-blocks compatible with positive semidefinite full correlation matrices. We consider 3{,}200 OOS realizations from the final walk-forward model with $n=300$, $\nu=0.3$, and $\Delta t_\textrm{in}=1000$. For each realization, we whiten the predicted block using Eq.~\eqref{eq:whitened_def} and record the canonical singular values. Figure~\ref{fig:CCC_boundedness} shows that feasibility violations are extremely rare in chronological data and absent in the shuffled control. The diagnostic also confirms that feasibility is not equivalent to unwhitened singular-value shrinkage: some modes may be amplified relative to their empirical values while the whitened canonical correlations remain within the admissible range.

\subsection{Tracking-error minimization as a differentiable module}
\label{app:tracking_error}
This appendix describes how the $\mathbf{C}_{XY}$ cleaning module is embedded in a broader end-to-end tracking-error architecture based on \textcite{bongiorno2025neural} and implemented with \textcite{rienet2025}. Let $\mathbf{R}_X\in\mathbb{R}^{n_x\times\Delta t_{\mathrm{in}}}$ and $\mathbf{R}_Y\in\mathbb{R}^{n_y\times\Delta t_{\mathrm{in}}}$ denote the in-sample return matrices for the investable and target universes. Given target weights $\mathbf{w}_Y$, the objective is to choose $\mathbf{w}_X$ so that $\mathbf{w}_X^\top\mathbf{r}^X_t$ tracks $\mathbf{w}_Y^\top\mathbf{r}^Y_t$. The tracking-error variance is
\begin{equation}
\mathrm{TEV}(\mathbf{w}_X\mid\mathbf{w}_Y)=\mathrm{Var}\!\left[\mathbf{w}_X^\top\mathbf{r}^X_t-\mathbf{w}_Y^\top\mathbf{r}^Y_t\right].
\end{equation}
The network first applies the lag-transformation and marginal-volatility modules inherited from RIEnet, then filters $\widehat{\boldsymbol{\Sigma}}_{XX}^{-1}$ and $\widehat{\boldsymbol{C}}_{XY}$. After rescaling the cleaned cross-correlation block, the unconstrained budget-only solution is
\begin{equation}
\mathbf{w}_X=\widetilde{\boldsymbol{\Sigma}}_{XX}^{-1}\widetilde{\boldsymbol{\Sigma}}_{XY}\mathbf{w}_Y-
\widetilde{\boldsymbol{\Sigma}}_{XX}^{-1}\mathbf{1}\,
\frac{\mathbf{1}^\top\widetilde{\boldsymbol{\Sigma}}_{XX}^{-1}\widetilde{\boldsymbol{\Sigma}}_{XY}\mathbf{w}_Y-1}{\mathbf{1}^\top\widetilde{\boldsymbol{\Sigma}}_{XX}^{-1}\mathbf{1}}.
\end{equation}
The training loss is the realized OOS tracking-error variance over the subsequent holding window. The configuration follows \ref{app:Training}, except that $\Delta t_{\mathrm{out}}=10$ trading days and the objective is the realized tracking error. This shorter horizon mitigates survivorship-related conditioning as discussed in \ref{app:delisting}.

\subsubsection{Tracking Error OOS Validation}
In validation, portfolio weights are obtained by solving the common long-only problem
\begin{equation}
\mathbf{w}_X^{\mathrm{val}}=\arg\min_{\mathbf{w}_X}\left\{\mathbf{w}_X^\top\widetilde{\boldsymbol{\Sigma}}_{XX}\mathbf{w}_X-2\mathbf{w}_X^\top\widetilde{\boldsymbol{\Sigma}}_{XY}\mathbf{w}_Y\right\},
\end{equation}
subject to $\mathbf{1}^\top\mathbf{w}_X^{\mathrm{val}}=1$ and $\mathbf{w}_X^{\mathrm{val}}\ge0$. The same validation protocol is applied to all benchmark estimators. For the sample benchmark, the covariance blocks are empirical. For the hybrid benchmarks, $\boldsymbol{\Sigma}_{XX}$ is cleaned with QIS and $\boldsymbol{\Sigma}_{XY}$ is obtained from either CV or BBP after rescaling the cleaned cross-correlation matrix. Each test year uses the model trained up to the previous year-end and evaluates 3{,}200 tracking instances with $(n_x,n_y,\Delta t_{\mathrm{in}})=(200,300,500)$ over the subsequent 10 trading days.

\section{Bootstrap Inference}
\label{app:bootstrap}

The significance statements in Tables~\ref{tab:synthetic_benchmarks},
\ref{tab:MSE_real}, \ref{tab:tracking_error}, and~\ref{tab:MSE_no_mode}
use paired comparisons of realized losses. Let $c$ and $c'$ index two
estimators, and let $\varphi_{c,\bullet}$ denote the realized loss of estimator
$c$, where $\bullet$ represents the relevant observation index. The paired
loss differential is
\begin{equation}
\psi_{c,c',\bullet}
:=
\varphi_{c,\bullet}-\varphi_{c',\bullet}.
\label{eq:paired_loss_differential}
\end{equation}
Positive values of $\psi_{c,c',\bullet}$ favor estimator $c'$. The loss
$\varphi_{c,\bullet}$ is the MSE in
Tables~\ref{tab:synthetic_benchmarks}, \ref{tab:MSE_real},
and~\ref{tab:MSE_no_mode}, and the realized OOS tracking error in
Table~\ref{tab:tracking_error}. The comparison follows the loss-differential
framework of \textcite{diebold1995comparing}, with a resampling scheme
adapted to the dependence structure of each experiment.

\subsection{Bootstrap for Synthetic Data}

For each row of Table~\ref{tab:synthetic_benchmarks}, all estimators are
evaluated on the same $n_{\mathrm{mc}}=1000$ independent Monte Carlo
realizations. Let $\iota=1,\ldots,n_{\mathrm{mc}}$ index these realizations.
The realization-level differential and its sample mean are
\begin{equation}
\psi_{c,c',\iota}
:=
\varphi_{c,\iota}-\varphi_{c',\iota},
\qquad
\widehat{\psi}_{c,c'}
:=
\frac{1}{n_{\mathrm{mc}}}
\sum_{\iota=1}^{n_{\mathrm{mc}}}
\psi_{c,c',\iota}.
\label{eq:synthetic_mean_differential}
\end{equation}

For bootstrap replication $\chi$, we draw
$\iota_{\upsilon}^{\ast(\chi)}$,
$\upsilon=1,\ldots,n_{\mathrm{mc}}$, independently with replacement from
$\{1,\ldots,n_{\mathrm{mc}}\}$ and compute
\begin{equation}
\widehat{\psi}_{c,c'}^{\ast(\chi)}
:=
\frac{1}{n_{\mathrm{mc}}}
\sum_{\upsilon=1}^{n_{\mathrm{mc}}}
\psi_{c,c',\iota_{\upsilon}^{\ast(\chi)}}.
\label{eq:synthetic_bootstrap_differential}
\end{equation}
The same resampled realization indices are used for all estimators, preserving
the pairing induced by their evaluation on a common simulated dataset. This
procedure is the ordinary nonparametric paired bootstrap
\parencite{efron1979bootstrap}. We use
$\beta_{\mathrm{boot}}=10{,}000$ bootstrap replications for each synthetic
condition.

\subsection{Bootstrap for Real Data}

For the real-data experiments, the resampling unit is the forecast date.
Let $t=1,\ldots,\tau$ index the forecast dates and let
$j=1,\ldots,m_t$ index the random asset partitions evaluated at date $t$.
For estimator $c$, we first average the realized loss across all partitions
associated with the same date,
\begin{equation}
\overline{\varphi}_{c,t}
:=
\frac{1}{m_t}
\sum_{j=1}^{m_t}
\varphi_{c,t,j}.
\label{eq:date_average_loss}
\end{equation}
The paired loss differential between estimators $c$ and $c'$ is then
\begin{equation}
\psi_{c,c',t}
:=
\overline{\varphi}_{c,t}
-
\overline{\varphi}_{c',t}.
\label{eq:date_loss_differential}
\end{equation}

For each bootstrap replication $\chi$, we draw
$\tau$ forecast dates with replacement from
$\{1,\ldots,\tau\}$. The same resampled dates are used for all
estimators, preserving the paired comparison. The bootstrap mean
differential is
\begin{equation}
\widehat{\psi}_{c,c'}^{\ast(\chi)}
:=
\frac{1}{\tau}
\sum_{\upsilon=1}^{\tau}
\psi_{c,c',t_{\upsilon}^{\ast(\chi)}}.
\label{eq:date_bootstrap_differential}
\end{equation}
We repeat this procedure $\beta_{\mathrm{boot}}=100{,}000$ times. For
year-specific results, dates are resampled within the corresponding test
year. For results aggregated over 2017-2024, the bootstrap mean is first
computed separately within each year and the eight annual means are then
combined with equal weight. Statistical significance is determined from the
resulting bootstrap distribution of the paired mean loss differential. This is
a paired nonparametric bootstrap at the forecast-date level
\parencite{efron1979bootstrap}.
  \printbibliography[title={Supplementary References}]

@inproceedings{troiani2022optimal,
  title={Optimal denoising of rotationally invariant rectangular matrices},
  author={Troiani, Emanuele and Erba, Vittorio and Krzakala, Florent and Maillard, Antoine and Zdeborov{\'a}, Lenka},
  booktitle={Mathematical and Scientific Machine Learning},
  pages={97--112},
  year={2022},
  organization={PMLR},
  doi={10.48550/arXiv.2203.07752}
}

@article{benaych2019optimal,
  title={Optimal cleaning for singular values of cross-covariance matrices},
  author={Benaych-Georges, Florent and Bouchaud, Jean-Philippe and Potters, Marc},
  journal={Ann. Appl. Probab.},
  volume={33},
  number={2},
  pages={1295--1326},
  year={2023},
  publisher={Institute of Mathematical Statistics},
  doi={10.1214/22-AAP1842}
}

@article{bongiorno2023filtering,
  title={Filtering time-dependent covariance matrices using time-independent eigenvalues},
  author={Bongiorno, Christian and Challet, Damien and Loeper, Gr{\'e}goire},
  journal={J. Stat. Mech. Theory Exp.},
  volume={2023},
  number={2},
  pages={023402},
  year={2023},
  publisher={IOP Publishing},
  doi={10.1088/1742-5468/acb7ed}
}

@article{bongiorno2025end,
  title={End-to-end large portfolio optimization for variance minimization with neural networks through covariance cleaning},
  author={Bongiorno, Christian and Manolakis, Efstratios and Mantegna, Rosario N.},
  journal={J. Financ. Data Sci.},
  volume={12},
  pages={100179},
  year={2026},
  issn={2405-9188},
  doi={10.1016/j.jfds.2026.100179},
  url={https://www.sciencedirect.com/science/article/pii/S2405918826000048}
}

@inproceedings{bongiorno2025neural,
  title={Neural Network-Driven Volatility Drag Mitigation under Aggressive Leverage},
  author={Bongiorno, Christian and Manolakis, Efstratios and Mantegna, Rosario N.},
  booktitle={Proceedings of the 6th ACM International Conference on AI in Finance},
  pages={449--455},
  year={2025},
  doi={10.1145/3768292.3770370}
}

@article{gavish2017optimal,
  title={Optimal shrinkage of singular values},
  author={Gavish, Matan and Donoho, David L.},
  journal={IEEE Trans. Inf. Theory},
  volume={63},
  number={4},
  pages={2137--2152},
  year={2017},
  publisher={IEEE},
  doi={10.1109/TIT.2017.2653801}
}

@article{stambaugh1999predictive,
  title={Predictive regressions},
  author={Stambaugh, Robert F.},
  journal={J. Financ. Econ.},
  volume={54},
  number={3},
  pages={375--421},
  year={1999},
  doi={10.1016/S0304-405X(99)00041-0}
}

@article{pitkajarvi2020cross,
  title={Cross-asset signals and time series momentum},
  author={Pitk{\"a}j{\"a}rvi, Aleksi and Suominen, Matti and Vaittinen, Lauri},
  journal={J. Financ. Econ.},
  volume={136},
  number={1},
  pages={63--85},
  year={2020},
  doi={10.1016/j.jfineco.2019.02.011}
}

@article{badrinath1995shepherds,
  title={Of shepherds, sheep, and the cross-autocorrelations in equity returns},
  author={Badrinath, Swaminathan G. and Kale, Jayant R. and Noe, Thomas H.},
  journal={Rev. Financ. Stud.},
  volume={8},
  number={2},
  pages={401--430},
  year={1995},
  doi={10.1093/rfs/8.2.401}
}

@article{hong2007industries,
  title={Do industries lead stock markets?},
  author={Hong, Harrison and Torous, Walter and Valkanov, Rossen},
  journal={J. Financ. Econ.},
  volume={83},
  number={2},
  pages={367--396},
  year={2007},
  doi={10.1016/j.jfineco.2005.09.010}
}

@article{awijen2023machine,
  title={Machine learning for US cross-industry return predictability under information uncertainty},
  author={Awijen, Haithem and Zaied, Younes B. and Lahouel, B{\'e}chir B. and Khlifi, Foued},
  journal={Res. Int. Bus. Finance},
  volume={64},
  pages={101893},
  year={2023},
  doi={10.1016/j.ribaf.2023.101893}
}

@article{bun2017cleaning,
  title={Cleaning large correlation matrices: tools from random matrix theory},
  author={Bun, Jo{\"e}l and Bouchaud, Jean-Philippe and Potters, Marc},
  journal={Phys. Rep.},
  volume={666},
  pages={1--109},
  year={2017},
  publisher={Elsevier},
  doi={10.1016/j.physrep.2016.10.005}
}

@article{chopra1993effect,
  title={The effect of errors in means, variances, and covariances on optimal portfolio choice},
  author={Chopra, Vijay K. and Ziemba, William T.},
  journal={J. Portf. Manag.},
  volume={19},
  number={2},
  pages={6--11},
  year={1993},
  publisher={World Scientific},
  doi={10.3905/jpm.1993.409440}
}

@book{muirhead2009aspects,
  title={Aspects of multivariate statistical theory},
  author={Muirhead, Robb J.},
  pages={239--246,585--586},
  year={2009},
  publisher={John Wiley \& Sons}
}

@article{ouellette1981schur,
  title={Schur complements and statistics},
  author={Ouellette, Diane V.},
  journal={Linear Algebra Appl.},
  volume={36},
  pages={187--295},
  year={1981},
  publisher={Elsevier},
  doi={10.1016/0024-3795(81)90232-9}
}

@article{carhart1997persistence,
  title={On persistence in mutual fund performance},
  author={Carhart, Mark M.},
  journal={J. Finance},
  volume={52},
  number={1},
  pages={57--82},
  year={1997},
  doi={10.1111/j.1540-6261.1997.tb03808.x}
}

@article{fama2015five,
  title={A five-factor asset pricing model},
  author={Fama, Eugene F. and French, Kenneth R.},
  journal={J. Financ. Econ.},
  volume={116},
  number={1},
  pages={1--22},
  year={2015},
  doi={10.1016/j.jfineco.2014.10.010}
}

@article{engle1982autoregressive,
  title={Autoregressive conditional heteroscedasticity with estimates of the variance of United Kingdom inflation},
  author={Engle, Robert F.},
  journal={Econometrica},
  volume={50},
  number={4},
  pages={987--1007},
  year={1982},
  doi={10.2307/1912773}
}

@article{bollerslev1986generalized,
  title={Generalized autoregressive conditional heteroskedasticity},
  author={Bollerslev, Tim},
  journal={J. Econom.},
  volume={31},
  number={3},
  pages={307--327},
  year={1986},
  doi={10.1016/0304-4076(86)90063-1}
}

@article{diebold1989factor,
  title={The dynamics of exchange rate volatility: A multivariate latent factor ARCH model},
  author={Diebold, Francis X. and Nerlove, Marc},
  journal={J. Appl. Econom.},
  volume={4},
  number={1},
  pages={1--21},
  year={1989},
  doi={10.1002/jae.3950040102}
}

@article{engle1990asset,
  title={Asset pricing with a factor-ARCH covariance structure: Empirical estimates for Treasury bills},
  author={Engle, Robert F. and Ng, Victor K. and Rothschild, Michael},
  journal={J. Econom.},
  volume={45},
  number={1--2},
  pages={213--237},
  year={1990},
  doi={10.1016/0304-4076(90)90099-F}
}

@article{engle2002dynamic,
  title={Dynamic conditional correlation: A simple class of multivariate generalized autoregressive conditional heteroskedasticity models},
  author={Engle, Robert F.},
  journal={J. Bus. Econ. Stat.},
  volume={20},
  number={3},
  pages={339--350},
  year={2002},
  doi={10.1198/073500102288618487}
}

@article{diebold1995comparing,
  title={Comparing predictive accuracy},
  author={Diebold, Francis X. and Mariano, Roberto S.},
  journal={J. Bus. Econ. Stat.},
  volume={13},
  number={3},
  pages={253--263},
  year={1995},
  doi={10.1080/07350015.1995.10524599}
}

@article{kelly2023principal,
  title={Principal portfolios},
  author={Kelly, Bryan T. and Malamud, Semyon and Pedersen, Lasse Heje},
  journal={J. Finance},
  volume={78},
  number={1},
  pages={347--387},
  year={2023},
  doi={10.1111/jofi.13199}
}

@article{cesarone2025benchmark,
  title={A benchmark-asset principal component factorization for index tracking on large investment universes},
  journal={Finance Research Letters},
  volume={79},
  pages={107244},
  year={2025},
  issn={1544-6123},
  doi={10.1016/j.frl.2025.107244},
  url={https://www.sciencedirect.com/science/article/pii/S1544612325005070},
  author={Cesarone, F. and {Di Paolo}, A. and Bufalo, M. and Orlando, G.}
}

@article{Lam2016NERCOME,
  author={Lam, Clifford},
  title={Nonparametric Eigenvalue-Regularized Precision or Covariance Matrix Estimator},
  journal={Ann. Stat.},
  year={2016},
  volume={44},
  number={3},
  pages={928--953},
  doi={10.1214/15-AOS1393}
}

@article{AbadirDistasoZikes2014,
  author={Abadir, Karim M. and Distaso, Walter and {\v Z}ike{\v s}, Filip},
  title={Design-free estimation of variance matrices},
  journal={J. Econom.},
  year={2014},
  volume={181},
  number={2},
  pages={165--180},
  doi={10.1016/j.jeconom.2014.03.010}
}

@article{ledoit2022quadratic,
  title={Quadratic shrinkage for large covariance matrices},
  author={Ledoit, Olivier and Wolf, Michael},
  journal={Bernoulli},
  volume={28},
  number={3},
  pages={1519--1547},
  year={2022},
  publisher={Bernoulli Society for Mathematical Statistics and Probability},
  doi={10.3150/20-BEJ1315}
}

@misc{sec12d22,
  title={{17 CFR \S 240.12d2-2 -- Removal from Listing and Registration}},
  author={{U.S. Securities and Exchange Commission}},
  howpublished={\url{https://www.law.cornell.edu/cfr/text/17/240.12d2-2}},
  year={2024},
  note={Accessed 20 April 2026}
}

@misc{rienet2025,
  title={RIEnet: A Rotational Invariant Estimator Network for Global Minimum-Variance Optimisation [software]},
  author={Bongiorno, Christian},
  year={2026},
  version={1.1.6},
  url={https://github.com/bongiornoc/RIEnet},
  note={Accessed 20 April 2026}
}

@misc{Opt_clean_BBP,
  title={Optimal cleaning for singular values of cross-covariance matrices [software]},
  author={Benaych-Georges, Florent},
  year={2021},
  url={https://github.com/CFMTech/Optimal_cleaning_for_singular_values_of_cross-covariance_matrices},
  note={Accessed 20 April 2026}
}

@article{fan2008high,
  title={High dimensional covariance matrix estimation using a factor model},
  author={Fan, Jianqing and Fan, Yingying and Lv, Jinchi},
  journal={Journal of Econometrics},
  volume={147},
  number={1},
  pages={186--197},
  year={2008},
  doi={10.1016/j.jeconom.2008.09.017}
}

@misc{kennethFrenchDataLibrary,
  author={{Kenneth R. French}},
  title={Data Library},
  year={2026},
  note={Accessed 20 April 2026},
  url={https://mba.tuck.dartmouth.edu/pages/faculty/ken.french/data_library.html}
}

@article{efron1979bootstrap,
  title={Bootstrap methods: Another look at the jackknife},
  author={Efron, Bradley},
  journal={Ann. Stat.},
  volume={7},
  number={1},
  pages={1--26},
  year={1979},
  doi={10.1214/aos/1176344552}
}

@misc{manolakis2026crienet,
  author       = {Bongiorno, Christian and Manolakis, Efstratios},
  title        = {{CRIENet}: {CrossRIEnet} for Rectangular
                  Cross-Correlation Cleaning},
  year         = {2026},
  howpublished = {\url{https://github.com/bongiornoc/CrossRIEnet}},
  note         = {Python software, version 0.2.0; accessed August 1, 2026}
}

\end{refsection}

\end{document}